\documentclass[aps,twocolumn,amsmath,amssymb,10pt]{revtex4}
\usepackage{hyperref}
\usepackage{graphicx}

\usepackage{color}
\def\d{\mathrm d}

\def\trace{\mathrm{tr}}
\def\ket#1{|#1\rangle}
\def\bra#1{\langle #1 |}

\begin{document}
\title{How two spins can thermalize a third spin}
\author{Stephan Kleinb\"olting and Rochus Klesse}
\affiliation{Universit\"at zu K\"oln, Institut f\"ur Theoretische Physik,
  Z\"ulpicher Str. 77, D-50937 K\"oln, Germany} \date{\today}
\begin{abstract}
We consider thermalization of a microscopic quantum system by interaction with a
thermal bath. Our interest is the minimal size the bath can have while
still being able to thermalize the system. Within a specific
thermalization scheme we show that a single spin-1/2 can be fully
thermalized by interaction with a bath that consists of just two other
spin-1/2. The two bath spins are initially in a pure, entangled state,
and the thermalizing interaction is a Heisenberg exchange-interaction
of the system-spin with one of the bath spins. The time dependent
coupling strength has to obey a single integral constraint. 
We also  present a simple generalization of this minimal model in
which the bath consists of an arbitrary number of spin-1/2 pairs.
\end{abstract}

\maketitle
Our topic is thermalization of a system by means of an interaction with
a bath, as specified in detail below. A bath that is suitable for that
purpose has to exhibit certain features. Particularly,
the standard textbook discussion explicitly or tacitly assumes that
some kind of thermodynamic limit regarding the bath has to be taken in
order to guarantee perfect thermalization of the system. The bath
is assumed to be {\em macroscopic} in a certain sense, also if the
system to be thermalized is a microscopic one.

The purpose of this brief note is to challenge the view that the
bath should always be a macroscopic one.
To this end we present a simple quantum-mechanical model in which the system
is a single spin-$1/2$, and the bath consists of only two other
spin-$1/2$. Initially, the two bath spins are in an entangled, pure
state.  
The system spin is shown to perfectly thermalize by means of a
finite-time exchange interaction with one of the bath spins.
As will be shown below, the mechanism at work is essentially swapping 
the states of system spin and one bath spin by the exchange interaction.
As a result, the entangled initial state of the two bath spins becomes the final 
entangled state of the system spin and the second bath spin. Tracing out the bath
spin yields the final thermal state of the system. 
This mechanism is neither new nor particularly intriguing. However, to the best of our knowledge, it
has not yet been discussed in the context of thermalization. 
Doing so may help to gain further insights in this problem: 
as simple as our model is, it very clearly points out the important role
entanglement can play in the thermalization of a system. 
We note that this aspect is crucial in modern theoretical approaches to
the phenomenon of thermalization in general \cite{thermalization}.

We will discuss {\em thermalization by a bath} within the following
general scheme. A closed quantum system $A$, the {\em system}, is
initially prepared in an arbitrary state 
associated with a density operator $\rho_A$. Aside $A$ there is another closed 
quantum system, the {\em bath} $B$.  Irrespective of the system's state
$\rho_A$, initially the bath is always in a certain state associated with a density operator $\rho_B$.
The third ingredient of the scheme is a time dependent interaction %$H_{AB}(t)$ 
that is supposed to establish thermal contact between system $A$ and bath
$B$ during a finite period of time $[0,T]$. Particularly, this means
that the interaction vanishes for $t<0$ and $t>T$. By the interaction
and the internal dynamics of $A$ and $B$, the initial
state $\rho_A \otimes \rho_B$ of the joint system at time $t=0$
evolves to a final joint state $\rho_{AB}'$ at time $t=T$.
We say that the scheme thermalizes system $A$ at some temperature
$\beta^{-1}$,
if for any initial state $\rho_A$ of $A$ the final reduced state of
$A$,
\begin{align*}
\rho_A'  = \trace_B \rho_{AB}',
\end{align*}
is the thermal state $\tau = e^{-\beta H_A}/Z$ of the canonical
ensemble. Here, $H_A$ is 
the Hamiltonian of system $A$ and $Z=\trace e^{-\beta H_A}$ is its partition sum.
  
The above scheme is idealistic as it requires the system to be fully
thermalized after an interaction during a finite period of time, $[0,T]$.
To obtain a more realistic scheme one could demand that $\rho'_A$
reaches the thermal state $\tau$ within some meaningful
approximation, if necessary, for $T \to \infty$. Here we
stay with the idealized scheme simply because our model meets its
stronger conditions.

Admittedly, our scheme uses the notion {\em bath} in a loose way.
For instance, it is not required that the bath itself is in 
some thermal equilibrium, we do not demand that it is able to 
thermalize an entire class of (microscopic) systems, and we also do
not insist that the bath is suitable for thermalizing the system by an
entire class of system-bath interactions.
Significantly strengthening the scheme in this point will probably
make it impossible to construct a microscopic bath. On the other hand,
we think that the presented scheme still captures essential features
of a general thermalization process.

It is instructive to view the thermalization of system $A$ formally as a
(rather simple) quantum operation ${\cal T}$ that maps an arbitrary
initial state $\rho_A$ 
onto the thermal state $\tau$. Being a proper quantum operation
\cite{Kra83,BZ08}, 
${\cal T}$ can be represented by an isometry $V:\cal H_A \to \cal H_A
\otimes \cal H_B$ that is followed by taking the partial trace with
respect to $\cal H_B$, i.e. ${\cal T}(\rho_A) = \trace_B V\rho_AV^\dagger$
\cite{Sti55,Cho75}. Here, $\cal H_A$ denotes the Hilbert space of
system $A$, and $\cal H_B$ is some appropriate ancilla Hilbert space.
Since ${\cal T}$ describes thermalization into $\tau$, we have 
\begin{align}\label{isometry}
\trace_B V\rho_A V^\dagger \: = \: \tau\:
\end{align}
for any initial state $\rho_A$. This relation can be phrased in more
physical terms if the isometry $V$ is expressed via an unitary
operator $U$ on $ \cal H_A\otimes \cal H_B$ and a normalized vector
$\ket \psi \in \cal H_B$ as 
\begin{align*}
V \ket\phi = U \ket \phi \otimes \ket \psi\:.
\end{align*}
Then, Eq. (\ref{isometry}) becomes 
\begin{align}\label{unitary}
\trace_B \: U  \rho_A \otimes \psi \,  U^\dagger = \tau\:,
\end{align}
where $\psi$ denotes the projection $\ket \psi \bra \psi$.
This relation might be interpreted within our thermalization scheme as
follows: $\cal H_B$ is the Hilbert space of a certain bath system $B$,
$\psi$ is its initial pure state, and $U$ is the joint 
quantum-mechanical time-evolution of $A$ and $B$.

Thus far, mathematical objects have been merely named with physical terms.
The question is whether there is a ${\cal T}$ representing isometry $V$ such that the 
associated $\cal H_B$, $U$ and $\psi$ have a physically meaningful interpretation. 
A quite simple answer to that question is suggested if one looks at
the dimensions $d_A$ and $d_B$ of the involved Hilbert spaces $\cal
H_A$ and $\cal H_B$. Assuming that $d_A$ is finite, 
it is known that $V$ can be realized as long as $d_B \ge
d_A^2$. Hence, it is possible to find an isometry $V$ for the choice $\cal H_B =
\cal H_A \otimes \cal H_A$, which means that in this case the bath $B$
can be thought to consist of just two copies $A_1$
and $A_2$ of the system $A_0\equiv A$. This, in turn, suggests
to chose the initial pure state $\ket \psi \in 
\cal H_A \otimes \cal H_A$ as a purification of the thermal state 
$\tau$ of $A_1$. I.e., for energy eigenstates 
$\ket 0_1, \ket 1_1,\dots, \ket{d_A-1}_1$
 of $A_1$ with energies $E_0, \dots E_{d_A-1}$,
and for arbitrary orthogonal states 
$\ket 0_2, \ket 1_2,\dots, \ket{d_A-1}_2$ of $A_2$ let 
\begin{align}\label{purification}
\ket \psi = \frac{1}{Z} \sum_k e^{-\beta E_k} \ket k_1 \otimes \ket k_2\:,
\end{align}
which clearly satisfies $\trace_{A_2} \psi = \tau$.
As a consequence of this choice the joint unitary time evolution
must be a product of a swap operation $W$ on $A_0$ and $A_1$, 
\begin{align}
W \ket k_0 \otimes \ket{ k'}_1 = \ket{ k'}_0 \otimes \ket{ k}_1\:,
\end{align}
and an arbitrary unitary operation $U_2$ on $A_2$,
\begin{align}\label{swap}
U = W \otimes U_2.
\end{align}
The effect on
$A_0$ in an arbitrary initial state $\rho_A$ is 
\begin{align*}
& \trace_B\: (W \otimes U_2)\: \rho_A \otimes \psi \: (W \otimes
  U)_2^\dagger \\
=\: & \trace_{A_1} W ( \rho_A \otimes \trace_{A_2} (\mathbf 1_1 \otimes
U_2) \psi
(\mathbf 1_1 \otimes U_2^\dagger) )W^\dagger \\
=\:& \trace_{A_1} W (\rho_{A_1} \otimes \trace_{A_2} \psi) W^\dagger \\
=\:& \trace_{A_1} W (\rho_{A_1} \otimes \tau) W^\dagger \\
=\:& \trace_{A_1} \tau \otimes \rho_A = \tau\:.
\end{align*}
This is the desired thermalization of $A_0$.

For the remaining task of finding a physically meaningful Hamiltonian
that generates the above dynamics we restrict ourself to the simple
case of system $A_0$ being a spin-$1/2$ in a magnetic field. It is a
well-known fact that then the swap $W$ can be generated by the
Heisenberg-exchange interaction $\vec S^0 \cdot \vec S^1$
with an appropriate time-dependent coupling $J(t)$.
[We use standard physics notation in units where $\hbar =1$, i.e. 
$\vec S = (S_x,S_y,S_z) \equiv
\frac{1}{2}(\sigma_1,\sigma_2,\sigma_3)$, $S_x^0\equiv S_x^0 \otimes
\mathbf 1^1 \otimes \mathbf 1^2$, etc.] 
In the end, this leads to a Hamiltonian
\begin{align}\label{hamiltonian}
H(t) = \epsilon S_z^0 + \epsilon S_z^1 \: + \: J(t) \vec S^0 \cdot
\vec S^1 \: + \: H_2\:,
\end{align}
where $\epsilon$ denotes the common spin-splitting energy for the spins $A_0$
and $A_1$, and $H_2$ is an arbitrary Hamiltonian for the second bath spin
$A_2$. Making use of the identity $W = 2 \vec S^0 \cdot \vec S^1 +
\mathbf{1}/2$ it is straight forward to check that 
the time evolution $U = T_{\gets} \exp -i \int_0^T
H(t) dt$ generated by $H(t)$ over the time period $[0,T]$ can be written as
\begin{align}
U = \left(\cos \alpha \:  + \: i W \sin \alpha \:
 \right) \otimes U_2\:,
\end{align}
where the phase $\alpha$ is given by
\begin{align}
\alpha = \frac{1}{2}\int_0^T J(t) \d t\:,
\end{align}
and the unitary $U_2$ describes the time evolution of the second bath
spin. Provided the time-dependent  
coupling $J(t)$ is such that the phase $\alpha$ is an odd multiple of
$\pi/2$, the entire time evolution $U$ is precisely
of the form Eq. (\ref{swap}) that we aimed for.

\begin{figure}
\begin{picture}(0,0)%
\includegraphics{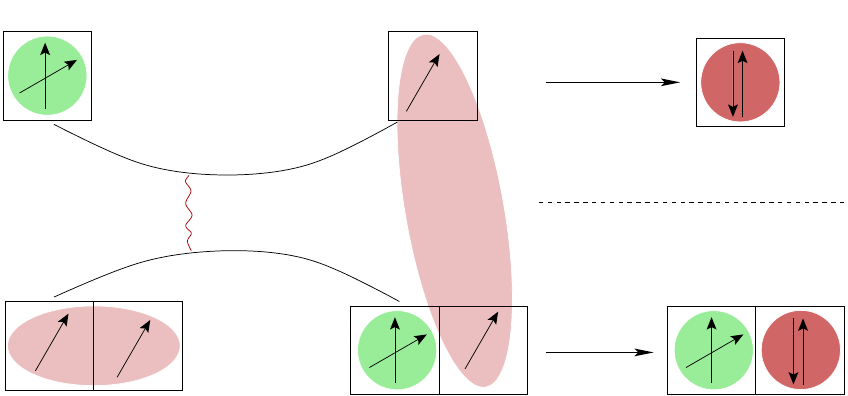}%
\end{picture}%
\setlength{\unitlength}{932sp}%
\begingroup\makeatletter\ifx\SetFigFont\undefined%
\gdef\SetFigFont#1#2#3#4#5{%
  \reset@font\fontsize{#1}{#2pt}%
  \fontfamily{#3}\fontseries{#4}\fontshape{#5}%
  \selectfont}%
\fi\endgroup%
\begin{picture}(17182,8029)(2191,-7328)
\put(3241,-1321){\makebox(0,0)[lb]{\smash{{\SetFigFont{9}{10.8}{\rmdefault}{\bfdefault}{\updefault}{\color[rgb]{0,0,0}$\rho$}%
}}}}
\put(3961,-6451){\makebox(0,0)[lb]{\smash{{\SetFigFont{9}{10.8}{\rmdefault}{\bfdefault}{\updefault}{\color[rgb]{0,0,0}$\psi$}%
}}}}
\put(16786,-6901){\makebox(0,0)[lb]{\smash{{\SetFigFont{9}{10.8}{\rmdefault}{\bfdefault}{\updefault}{\color[rgb]{0,0,0}$\rho$}%
}}}}
\put(18556,-6896){\makebox(0,0)[lb]{\smash{{\SetFigFont{9}{10.8}{\rmdefault}{\bfdefault}{\updefault}{\color[rgb]{0,0,0}$\tau$}%
}}}}
\put(17326,-1456){\makebox(0,0)[lb]{\smash{{\SetFigFont{9}{10.8}{\rmdefault}{\bfdefault}{\updefault}{\color[rgb]{0,0,0}$\tau$}%
}}}}
\put(6211,-3841){\makebox(0,0)[lb]{\smash{{\SetFigFont{9}{10.8}{\rmdefault}{\bfdefault}{\updefault}{\color[rgb]{0,0,0}$J(t) \vec{S}^1 \!\cdot\! \vec{S}^2$}%
}}}}
\put(2296,-5191){\makebox(0,0)[lb]{\smash{{\SetFigFont{9}{10.8}{\familydefault}{\mddefault}{\updefault}{\color[rgb]{0,0,0}$B$:}%
}}}}
\put(2206,254){\makebox(0,0)[lb]{\smash{{\SetFigFont{9}{10.8}{\familydefault}{\mddefault}{\updefault}{\color[rgb]{0,0,0}$A$:}%
}}}}
\put(11116,-3166){\makebox(0,0)[lb]{\smash{{\SetFigFont{9}{10.8}{\rmdefault}{\bfdefault}{\updefault}{\color[rgb]{0,0,0}$\psi$}%
}}}}
\put(10351,-6901){\makebox(0,0)[lb]{\smash{{\SetFigFont{9}{10.8}{\rmdefault}{\bfdefault}{\updefault}{\color[rgb]{0,0,0}$\rho$}%
}}}}
\put(13501,-871){\makebox(0,0)[lb]{\smash{{\SetFigFont{9}{10.8}{\rmdefault}{\bfdefault}{\updefault}{\color[rgb]{0,0,0}$\trace_B$}%
}}}}
\put(13501,-6361){\makebox(0,0)[lb]{\smash{{\SetFigFont{9}{10.8}{\rmdefault}{\bfdefault}{\updefault}{\color[rgb]{0,0,0}$\trace_A$}%
}}}}
\end{picture}%
 \caption{\label{fig-model} A system spin ($A$) in an arbitrary initial state
   $\rho$ interacts with two bath spins ($B$) in a pure, entangled initial state
   $\psi$. By the interaction the entanglement is swapped to the upper spin 
   and the reduced state of the upper spin becomes a thermal state $\tau$.}
 \end{figure}
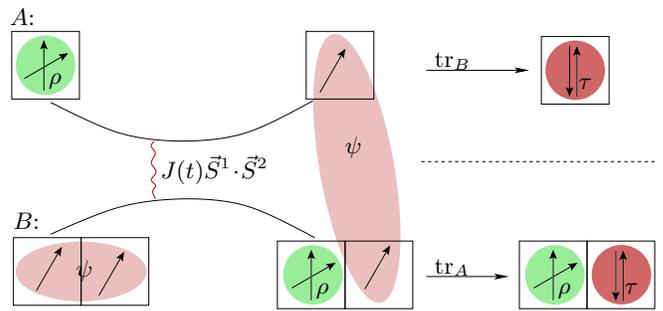

The minimal thermalization model thus obtained is summarized in Fig.~\ref{fig-model}. 
It is worth emphasizing that the two bath spins in the pure state
$\psi$ together with the proposed interaction always lead to complete
thermalization of the system spin, irrespective of the initial state
$\rho_A$ of the system spin. Also, the interaction is physical in the
sense that it is a simple two-spin interaction; a three-spin
interaction would comply with the mathematical formalism as well but is
actually not required.
Owing to the fact that the bath is microscopic its
dynamics has to match the dynamics of the system to be thermalized. 
For this reason the spin-splitting energy of the system spin must
equal the one of the first bath spin. In contrast to that, the
local dynamics of the second bath spin generated by $H_2$ does not change
the entanglement with the two other spins and therefore is 
irrelevant for the thermalization. This offers a trivial
interpretation of our model: Since the second bath spin is actually not
involved in the system-bath interaction it can be traced out right
from beginning. This leaves us with the first bath spin in the thermal
state $\tau$, which by the exchange interaction is then simply transferred
to the system spin.

Another important point is that the coupling $J(t)$ has to be chosen
such that the associated phase $\alpha$ must be an odd multiple of
$\pi/2$. Here it is to mention that also  
in case of a macroscopic bath an interaction that is able to
thermalize will have to obey conditions. For instance, 
a system-bath coupling that is instantaneously switched off will 
almost never leave the system in a thermal state.
Yet, it seems to be clear that a macroscopic bath allows for more
freedom in the choice of a thermalizing interaction. To illustrate
the latter point we briefly discuss a simple generalization
of the above minimal model.

The generalized model consists of a system spin $A_0$ and $n$ pairs of
bath spins denoted by $A_{\nu 1}$ and $A_{\nu 2}$ with $\nu=1,\dots,n$.
Initially, each spin pair $A_{\nu 1} A_{\nu 2}$ of the bath is in the
kind of pure state 
Eq. (\ref{purification}) as in the minimal model. This
defines an initial bath state $\psi_n$ as a product of $n$ entangled
two-spin states, each being a purification of $\tau$. Corresponding
to the Hamiltonian of the minimal model, Eq. (\ref{hamiltonian}),
the generalized model has the Hamiltonian
\begin{align*}
H=\epsilon S^0_z + \epsilon \sum_{\nu=1}^n S^{\nu 1}_z +
\sum_{\nu=1}^n J_\nu(t) \vec S^0 \cdot \vec S^{\nu 1} +
\sum_{\nu=1}^n H^{\nu 2}\:. 
\end{align*}
The exchange interactions $\vec S^0 \cdot \vec S^{\nu 1} $ do not
commute among each other. To allow for an analytical treatment we
therefore assume that the couplings $J_1(t), \dots, J_n(t)$ are supported on
disjoint time intervals $[0,t_1[, [t_1,t_2[, \dots, [t_n,T]$. With
this assumption the time evolution operator on $A_0, A_{11}, \dots,
A_{n1}$ can be written as the ordered product
\begin{align}
\prod^{\gets}_\nu (\cos \alpha_\nu + i W_{0\nu} \sin \alpha_\nu)\:,
\end{align}
where $W_{0\nu}$ is the swap operation on $A_0,A_{\nu 1}$, and $\alpha_\nu
= \frac{1}{2} \int_0^T J_\nu(t) \d t$. The effect of the entire time evolution 
on the system spin $A_0$ can be easily computed if first all spins
$A_{12},\dots, A_{n2}$ are traced out  and then successively the
partial traces over the bath spins $A_{11}, A_{21}, \dots, A_{n1}$ are
taken. This is conveniently done by representing the density operators 
as matrices w.r.t.\ to the energy eigenstates $\ket 0,
\ket 1$. In this way we find that an initial state of the system spin
$A_0$  
\begin{align}
\rho_0 = 
\begin{pmatrix} 
a_{00} & a_{01} \\
a_{10} & a_{11} \\
\end{pmatrix}
\end{align} 
becomes a final state 
\begin{align}
  \rho_0' = 
\begin{pmatrix} 
p \: a_{00} & \lambda \: a_{01} \\
\lambda^* \: a_{10} & p\:  a_{11} \\
\end{pmatrix}
 + (1-p) \tau,
\end{align}
where $\tau$ is the thermal state, and the parameters $p$ and
$\lambda$ are given by
\begin{align*}
p  = \prod_{\nu=1}^n  p_\nu\:, \quad \lambda = \prod_{\nu =1}^n
\lambda_\nu \:,
\end{align*}
with 
\begin{align*}
p_\nu &= \cos^2 \alpha_\nu\:, \\
\lambda_\nu &= \cos^2 \alpha_\nu \: + \: i \cos \alpha_\nu \sin
\alpha_\nu \tanh \frac{\beta \epsilon}{2}\:.
\end{align*}
The moduli of $p_\nu$ and $\lambda_\nu$ are bounded above by unity,
and they are less than unity as long as the phase  $\alpha_\nu$ is off
an even multiple of $\pi/2$. For this reason, in the large $n$-limit
generic interactions $J_\nu(t)$ will lead to very small parameters $p
\ll 1$ and $|\lambda| \ll 1$, meaning that the 
final state is very close to the thermal state $\tau$. This can be
made more precise by stating that the trace distance 
of $\rho_0'$ to $\tau$ is upper bounded by 
$ \sqrt{ p^2 + |\lambda|^2}$.
We conclude that the constraints that have to imposed onto the couplings
$J_\nu(t)$ in order to achieve thermalization are fairly weak in the
generalized model.

Finally, a few remarks on entropy in the minimal model. Initially, the
von Neumann entropy of the total system equals the entropy of
system $A$, $S(\rho_A)$, because the initial pure 
state $\psi$ of the bath has vanishing entropy. The unitary
joint time evolution $U$ leaves the total entropy unchanged, meaning
that the final state $\rho_{AB}'= U \rho_A \otimes \psi U^\dagger$
of the joint system has also entropy $S(\rho_A)$.
However, in the end the  total system is ``split'' into
system and bath. This means, by definition, that after the split
system and bath can be accessed only by local observables of the form 
$O_A \otimes \mathbf 1_B$ and $\mathbf 1_A \otimes O_B$. 
With respect to these local observables the final state $\rho_{AB}'$
is equivalent to a product state $\rho_A' \otimes
\rho_B'$ \cite{equivalence}, where the factors are the reduced states 
$\rho_A' = \trace_B \rho_{AB}' \equiv \tau$ and   
$\rho_B' = \trace_A \rho_{AB}' \equiv \rho_A \otimes U_2 \tau U_2^\dagger$.
Hence, the final entropy of the split system in state $\rho_A' \otimes
\rho_B'$ is  $2 S(\tau) + S(\rho_A)$, which exceeds the initial
entropy by $2S(\tau)$. This entropy increase is
precisely the entanglement entropy of the state $\rho_{AB}'$ 
w.r.t.\ to $A$ and $B$. In this way, the increase of the total entropy
is seen to be the effect of the separation of the total system
in system and bath.

Financial support by DFG grant ZI-513/1-2, by the center for
Quantum Matter and Materials (QM2) of the University
of Cologne, and by the SFB/TR 12 is gratefully acknowledged.


\begin{thebibliography}{10}
\bibitem{equivalence} Here, the equivalence of $\rho_{AB}'$ and
  $\rho_A' \otimes \rho_B'$ means that for any local operators 
  $O_A$, $O_B$ \ $\trace\: O_A \otimes   O_B \rho_{AB}' =
  \trace\: O_A \rho_A' \: \trace\:
 O_B \rho_B'$.
\end{thebibliography}
\end{document}